\documentclass[aps,prl,twocolumn,superscriptaddress]{revtex4-1} 

\usepackage{epsfig}
\usepackage{color}
\usepackage[latin1]{inputenc}
\usepackage{float,amsmath}
\usepackage{graphicx}

\newcommand{\be}{\begin{eqnarray}}
\newcommand{\ee}{\end{eqnarray}}

\newcommand{\benum}{\begin{enumerate}}
\newcommand{\eenum}{\end{enumerate}}

\begin{document}

\title{Event-by-event anisotropic flow in heavy-ion collisions\\ from combined Yang-Mills and viscous fluid dynamics}

\author{Charles Gale}
\affiliation{Department of Physics, McGill University, 3600 University Street, Montreal, Quebec, H3A\,2T8, Canada}
\author{Sangyong Jeon}
\affiliation{Department of Physics, McGill University, 3600 University Street, Montreal, Quebec, H3A\,2T8, Canada}
\author{Bj\"orn Schenke}
\affiliation{Physics Department, Brookhaven National Laboratory, Upton, NY 11973, USA}
\author{Prithwish Tribedy}
\affiliation{Variable Energy Cyclotron Centre, 1/AF Bidhan Nagar, Kolkata 700064, India}
\author{Raju Venugopalan}
\affiliation{Physics Department, Brookhaven National Laboratory, Upton, NY 11973, USA}

\begin{abstract}
Anisotropic flow coefficients $v_1$-$v_5$ in heavy ion collisions are computed by combining a classical Yang-Mills description of the early time glasma 
flow with the subsequent relativistic viscous hydrodynamic evolution of matter through the quark-gluon plasma and hadron gas phases. 
The glasma dynamics, as realized in the IP-Glasma model, takes into account event-by-event geometric fluctuations in nucleon positions and intrinsic sub-nucleon scale 
color charge fluctuations; the pre-equilibrium flow of matter is then matched to the \textsc{music} algorithm describing viscous hydrodynamic flow and particle production at freeze-out. 
The IP-Glasma+\textsc{music} model describes well both transverse momentum dependent and integrated $v_n$ data measured at the Large Hadron Collider (LHC) and the Relativistic Heavy Ion Collider (RHIC). The model also reproduces the event-by-event distributions of $v_2$, $v_3$ and $v_4$ measured by the ATLAS collaboration. 
The implications of our results for better understanding of the dynamics of the glasma as well as for the extraction of transport properties of the quark-gluon plasma are outlined. 
\end{abstract}

\maketitle


Heavy ion collisions at the Relativistic Heavy Ion Collider (RHIC) and the Large Hadron Collider (LHC) uniquely allow for systematic exploration of the high temperature many-body dynamics of a non-Abelian quantum field theory. Particularly intriguing is the prospect of disentangling the non-equilibrium strongly correlated dynamics of the early time glasma regime from those of late stage nearly equilibrated quark-gluon plasma and hadron gas phases by measurements of  anisotropic flow harmonics $v_n$ at both RHIC \cite{Adare:2011tg,Sorensen:2011fb} and LHC 
\cite{ALICE:2011ab,ATLAS:2012at,CMS:2011}.

An excellent candidate for providing initial conditions for systematic flow studies is the IP-Glasma model described in detail in Refs.~\cite{Schenke:2012wb,Schenke:2012hg}.  It combines the IP-Sat (Impact Parameter Saturation Model) model~\cite{Bartels:2002cj,Kowalski:2003hm} of high energy nucleon (and nuclear) wavefunctions with the classical Yang-Mills (CYM) dynamics of the glasma fields produced in a heavy-ion collision~\cite{Kovner:1995ja,Kovchegov:1997ke,Krasnitz:1998ns,*Krasnitz:1999wc,*Krasnitz:2000gz,Lappi:2003bi}. We note that the IP-Sat model provides a good description of small $x$ HERA deeply inelastic scattering (DIS) data off protons and fixed target nuclear DIS data~\cite{Kowalski:2006hc,*Kowalski:2007rw}. Prior implementation of the IP-Sat model in proton-proton and nucleus-nucleus collisions at the LHC using a $k_\perp$-factorized expression approximating CYM dynamics was shown to give good agreement with bulk features of data \cite{Tribedy:2010ab,*Tribedy:2011aa}. The upcoming p+Pb run at the LHC should provide further constraints on the dynamics of the IP-Glasma model, in particular the energy dependence of $Q_s$. 

In this letter, we couple the IP-Glasma model of the classical early time evolution of boost-invariant configurations of gluon fields to a relativistic
hydrodynamic description of the system, using the energy density and flow velocity in the transverse plane at the switching time $\tau_{\rm switch}\sim 1/Q_s$ as input. The hydrodynamic
evolution in each event is described by \textsc{music} \cite{Schenke:2010nt,Schenke:2010rr,Schenke:2011tv,Schenke:2011bn}, a 3+1 dimensional relativistic viscous hydrodynamic simulation~\cite{Israel:1979wp} that uses the Kurganov-Tadmor algorithm \cite{Kurganov:2000}. While this matching of glasma dynamics to viscous hydrodynamics is a significant improvement relative to previously employed initial conditions for heavy ion collisions, early stage dynamics is not fully included. Most notably, the hydrodynamic viscous tensor $\Pi^{\mu\nu}$ is too large to be described self-consistently by a gradient expansion. Instabilities triggered by  quantum fluctuations, and subsequent strong scattering of over-occupied fields, may lead to rapid quenching of  $\Pi^{\mu\nu}$ to reasonable values justifying the use of viscous hydrodynamics already at early times.  In this letter, we will assume such an efficient mechanism to be at work and set the initial value of $\Pi^{\mu\nu}$ to zero. Recent progress in computing early-time quantum fluctuations will help eliminate this systematic uncertainty~\cite{Dusling:2011rz,Dusling:2010rm,Epelbaum:2011pc,Dusling:2012ig2,Berges:2012mc}.

When we switch from the CYM description to hydrodynamics we construct the fluid's initial energy momentum tensor $T^{\mu\nu}_{\rm fluid} = (\epsilon + {\cal P})u^\mu u^\nu - {\cal P}g^{\mu\nu} + \Pi^{\mu\nu}$ from the energy density in the fluid's rest frame $\varepsilon$, the flow velocity $u^\mu$, and, using an equation of state, the local pressure ${\cal P}$ at each transverse position. $\varepsilon$ and $u^\mu$ are obtained by solving $u_\mu T^{\mu\nu}_{\rm CYM} = \varepsilon u^\nu$, using the fact that $u^\mu$ is a time-like eigenvector of $T^{\mu\nu}_{\rm CYM}$ and satisfies $u^2 =1$.  

Other important details of our analysis are as follows. Unless otherwise noted, $\tau_{\rm switch} = 0.2\,{\rm fm}/c$. 
We employ the \emph{s95p-PCE} equation of state, obtained from fits to lattice QCD results and a hadron resonance gas model \cite{Huovinen:2009yb},
with partial chemical equilibrium (PCE) setting in below a temperature $T_{\rm PCE}=150\,{\rm MeV}$. 
Kinetic freeze-out occurs at $T_{\rm FO}=120\,{\rm MeV}$. At this temperature, we implement the Cooper-Frye prescription \cite{Cooper:1974mv} for computing particle spectra.
Unless otherwise noted, shown results include decays from resonances of masses up to $1.3\,{\rm GeV}$.

\begin{figure}[tb]
   \begin{center}
     \includegraphics[width=8.75cm]{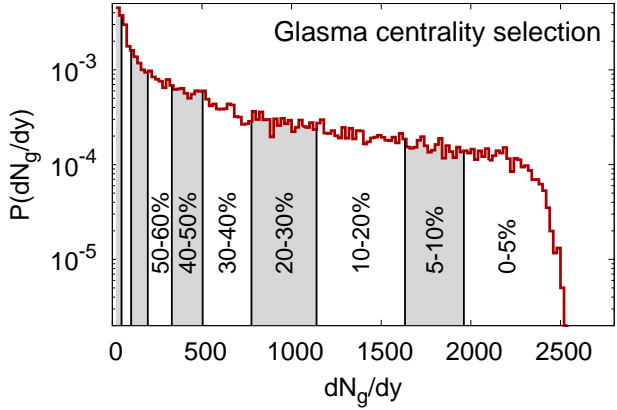}
     \vspace{-0.75cm}
     \caption{(Color online) Gluon multiplicity distribution in the IP-Glasma model.}
     \label{fig:dNdyCentrality}
   \end{center}
   \vspace{-0.5cm}
\end{figure}

A novel feature of our study is the determination of centrality classes using the multiplicity distribution of gluons much alike the procedure followed by the heavy ion experiments~\footnote{Strictly, this requires computing $10^4$ complete events to determine charged particle distributions. Because this is computationally very demanding, we instead determine centrality from the distribution of produced gluons. We showed previously this gives good agreement with the uncorrected distribution in the STAR experiment \cite{Schenke:2012hg}.}. The gluon multiplicity distribution is shown in Fig.\,\ref{fig:dNdyCentrality}. Centrality classes are determined from the fraction of the integral over this distribution, beginning with integrating from  the right. As a consequence of implementing this centrality selection, we properly account for impact parameter and multiplicity fluctuations.

\begin{figure}[tb]
   \begin{center}
     \includegraphics[width=8.75cm]{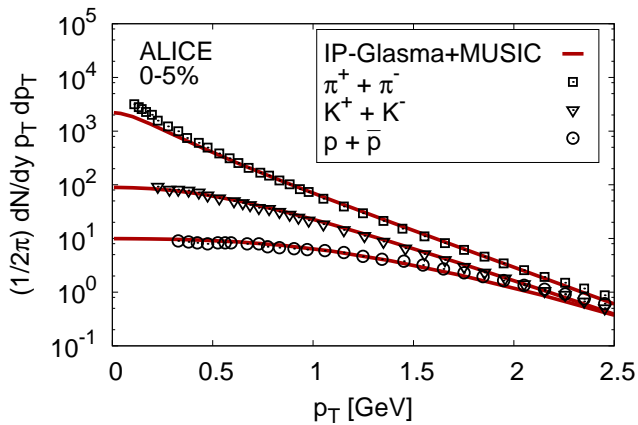}
     \vspace{-0.75cm}
     \caption{(Color online) Identified particle transverse momentum spectra including all resonances up to $2\, {\rm GeV}$ 
       compared to experimental data from the ALICE collaboration \cite{ALICE:2012iu}.}
     \label{fig:pt-}
   \end{center}
   \vspace{-0.5cm}
\end{figure}

Because entropy is produced during the viscous hydrodynamic evolution, we need to adjust the normalization of the initial energy density commensurately to describe the final particle spectra \footnote{A more detailed study of the effect of viscosity on particle production relative to that in the glasma stage will be reported separately.}. The obtained $p_T$-spectra of pions, kaons, and protons are shown for 0-5\% central collisions at $\sqrt{s}=2.76\,{\rm TeV}$/nucleon, using $\eta/s=0.2$, in Fig.\,\ref{fig:pt-}, and compared to data from ALICE \cite{ALICE:2012iu}. The results are for averages over only 20 events in this case, but statistical errors are smaller than the line width for the spectra.
Overall, the agreement with experimental data is good. However, soft pions at $p_T<300\,{\rm MeV}$ are underestimated. 

We determine $v_1$ to $v_5$ in every event by first determining the \emph{exact} event plane \footnote{We emphasize we are not using the experimental event-plane method. We compute a smooth average over particle distributions and there is no need for a resolution correction factor.}
\begin{equation}
  \psi_n=\frac{1}{n}\arctan\frac{\langle \sin(n\phi)\rangle}{\langle \cos(n\phi)\rangle}\,,
\end{equation}
and then computing 
\begin{align}
  v_n(p_T) &=\langle \cos(n(\phi-\psi_n)) \rangle \nonumber\\
  &\equiv {\int d\phi f(p_\perp,\phi) \cos(n(\phi-\psi_n))\over \int d\phi f(p_\perp,\phi)}\,,
\end{align}
where $f(p_\perp,\phi)$ are the thermal distribution functions obtained in the Cooper-Frye approach (with additional contributions from resonance decays).

\begin{figure}[tb]
   \begin{center}
     \includegraphics[width=8.75cm]{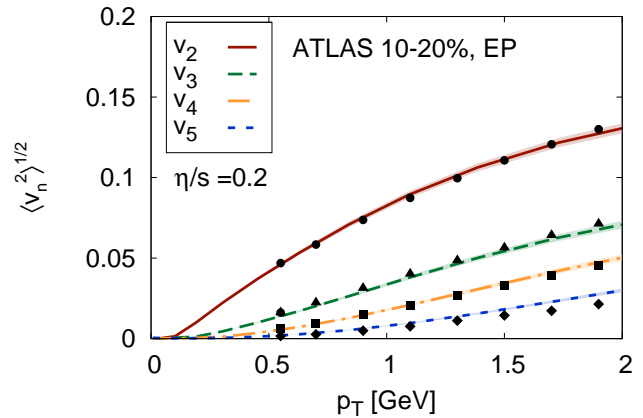}
     \vspace{-0.75cm}
     \caption{(Color online) Root-mean-square anisotropic flow coefficients $\langle v_n^2 \rangle ^{1/2}$ as a function of transverse momentum, 
       compared to experimental data by the ATLAS collaboration using the event plane (EP) method \cite{ATLAS:2012at} (points). 200 events. Bands indicate statistical errors. Experimental error bars are smaller than the size of the points.}
     \label{fig:vn10-20}
   \end{center}
   \vspace{-0.5cm}
\end{figure}

\begin{figure}[tb]
   \begin{center}
     \includegraphics[width=8.75cm]{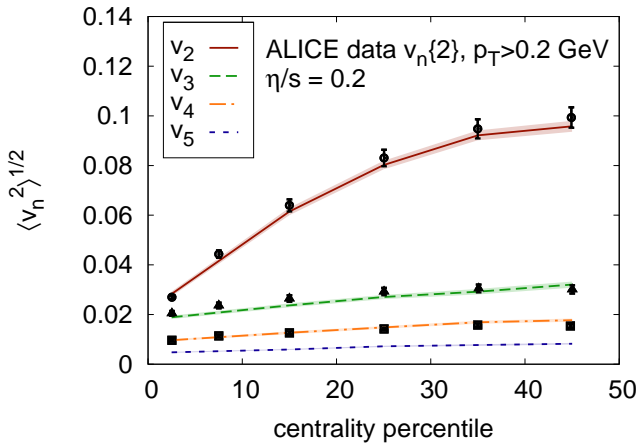}
     \vspace{-0.75cm}
     \caption{(Color online) Root-mean-square anisotropic flow coefficients $\langle v_n^2 \rangle ^{1/2}$, computed as a function of centrality, 
       compared to experimental data of $v_n\{2\}$, $n\in\{2,3,4\}$, by the ALICE collaboration \cite{ALICE:2011ab} (points). Results are for 200 events per centrality with bands indicating statistical errors.}
     \label{fig:vnCent}
   \end{center}
   \vspace{-0.5cm}
\end{figure}

We first present  the root-mean-square (rms) $v_n(p_T)$ for $10-20\%$ central collisions and compare to experimental data from the ATLAS collaboration \cite{ATLAS:2012at} in Fig.\,\ref{fig:vn10-20}. Agreement for $v_2$-$v_5$ is excellent. We note that the $v_n$ from the experimental event plane method do not exactly correspond to the rms values, but lie somewhere between the mean and the rms values. In this regard, a better comparison is the $p_T$-integrated rms $v_n$ to the ALICE $v_n\{2\}$ results--which correspond to the rms values. Excellent agreement over the whole studied centrality range is achieved for the experimentally available $v_2$, $v_3$ and $v_4$, as shown in Fig.\,\ref{fig:vnCent}.

We studied the effect of initial transverse flow included in our framework by also computing $v_n(p_T)$ with $u^\mu$ set to zero at time $\tau_{\rm switch}$.
The effect on hadron anisotropic flow turns out to be extremely weak - results agree within statistical errors. 
Because photons are produced early on in the collision, we expect a greater effect on photon anisotropic flow; this will be examined in a subsequent work. 
We emphasize that pre-equilibrium dynamics that is not fully accounted for may still influence the amount of initial transverse flow. 

\begin{figure}[tb]
   \begin{center}
     \includegraphics[width=8.75cm]{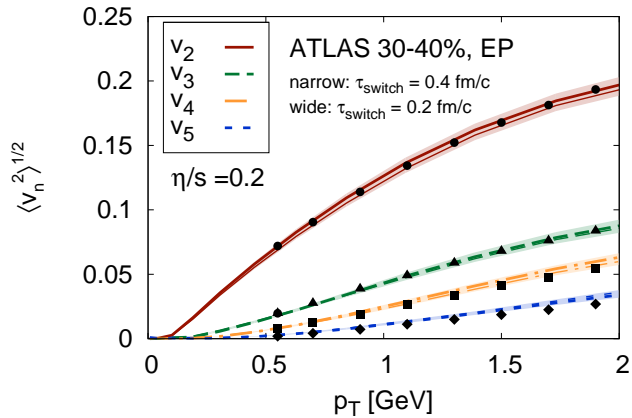}
     \vspace{-0.75cm}
     \caption{(Color online) Comparison of $v_n(p_T)$ using two different switching times $\tau_{\rm switch}=0.2\,{\rm fm}/c$ (wide), and $0.4\,{\rm fm}/c$
       (narrow). Experimental data by the ATLAS collaboration using the event-plane (EP) method \cite{ATLAS:2012at} (points). Bands indicate statistical errors. }
     \label{fig:vn30-40-switch}
   \end{center}
   \vspace{-0.5cm}
\end{figure}

The effect of changing the switching time from $\tau_{\rm switch}=0.2\,{\rm fm}/c$ to $\tau_{\rm switch}=0.4\,{\rm fm}/c$ is shown in Fig.\,\ref{fig:vn30-40-switch}.
Results agree within statistical errors, but tend to be slightly lower for the later switching time.
The nonlinear interactions of classical fields become weaker as the system expands and therefore Yang-Mills dynamics is 
less effective than hydrodynamics in building up flow at late times. Yet it is reassuring that there is a window in time where both
descriptions produce equivalent results.

\begin{figure}[tb]
   \begin{center}
     \includegraphics[width=8.75cm]{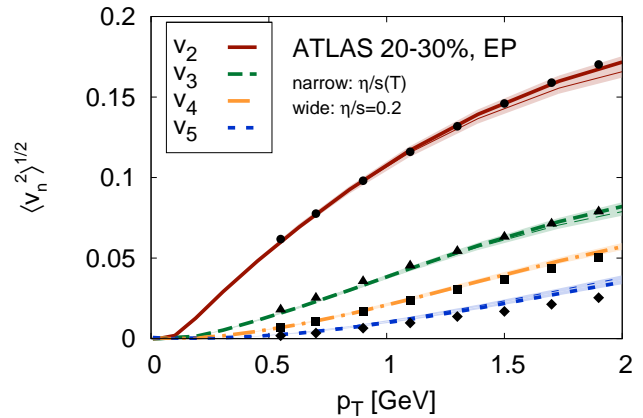}
     \vspace{-0.75cm}
     \caption{(Color online) Comparison of $v_n(p_T)$ using constant $\eta/s=0.2$ and a temperature dependent $\eta/s(T)$ as parametrized in \cite{Niemi:2011ix}. Experimental data by the ATLAS collaboration using the event-plane (EP) method \cite{ATLAS:2012at} (points).  Bands indicate statistical errors. }
     \label{fig:vn20-30-vTdep}
   \end{center}
   \vspace{-0.5cm}
\end{figure}

Because a constant $\eta/s$ is at best a rough effective measure of the evolving shear viscosity to entropy density ratio, we present 
results for a parametrized temperature dependent $\eta/s$, following \cite{Niemi:2011ix}.
We use the same parametrization (HH-HQ) as in \cite{Niemi:2011ix,Niemi:2012ry} with a minimum of $\eta/s(T)=0.08$ at $T=T_{\rm tr}=180\,{\rm MeV}$. The result, compared to $\eta/s=0.2$ is shown for $20-30\%$ central collisions in Fig.\,\ref{fig:vn20-30-vTdep}.
The results are indistinguishable when studying just one collision energy. The insensitivity of our results to two very different functional forms may suggest that a very large fraction of the magnitude of the flow coefficients is built up at later times when $\eta/s$ is very small. 
Also, since second order viscous hydrodynamics breaks down when $\Pi^{\mu\nu}$ is comparable to the ideal terms, our framework may be inadequate for large values of $\eta/s$.

\begin{figure}[tb]
   \begin{center}
     \includegraphics[width=8.75cm]{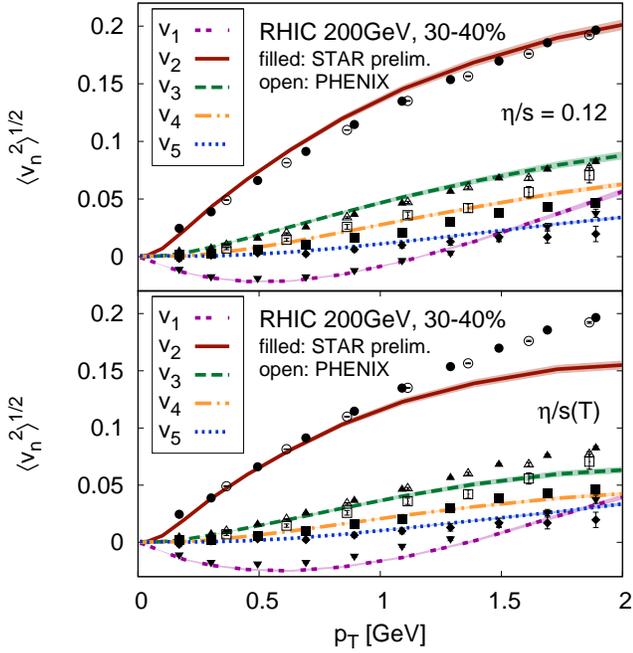}
     \vspace{-0.75cm}
     \caption{(Color online) Comparison of $v_n(p_T)$ at RHIC using constant $\eta/s=0.12$ and a temperature dependent $\eta/s(T)$ as parametrized in \cite{Niemi:2011ix}. Experimental data by the PHENIX \cite{Adare:2011tg} (open symbols) and STAR \cite{STAR:2012QM} (preliminary, filled symbols) collaborations. Bands indicate statistical errors. }
     \label{fig:vn30-40-RHIC}
   \end{center}
   \vspace{-0.5cm}
\end{figure}

At top RHIC energy, as shown in  Fig.\,\ref{fig:vn30-40-RHIC}, the experimental data from STAR \cite{STAR:2012QM} and PHENIX \cite{Adare:2011tg} is well described when using a constant $\eta/s=0.12$, which is about $40\,\%$ smaller than the value at LHC. A larger effective $\eta/s$ at LHC than at RHIC was also found in \cite{Song:2011qa}.
The temperature dependent $\eta/s(T)$ used to describe LHC data works well for low-$p_T$ RHIC data, but underestimates $v_2(p_T)$ and $v_3(p_T)$ for $p_T>1\,{\rm GeV}$.  
The parametrizations of $\eta/s(T)$ in the literature are not definitive and significant improvements are necessary. Our studies suggest great potential for extracting the temperature dependent properties of QCD transport coefficients by performing complementary experiments extracting flow harmonics at both RHIC and LHC.

\begin{figure}[tb]
   \begin{center}
     \includegraphics[width=8.75cm]{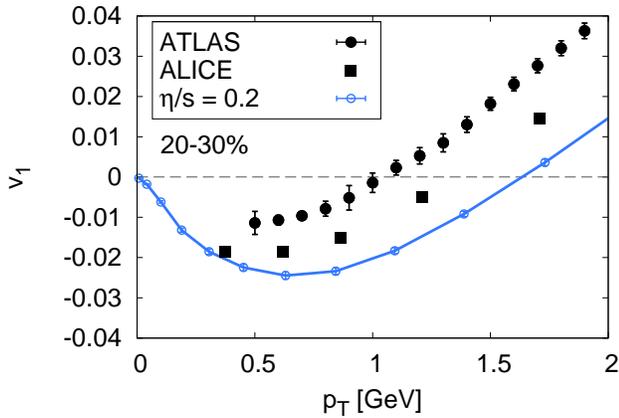}
     \vspace{-0.75cm}
     \caption{(Color online) $v_1(p_T)$ compared to experimental data from the ALICE \cite{Aamodt:2011by} and ATLAS \cite{Jia:2012hx} collaborations.}
     \label{fig:v1-20-30}
   \end{center}
   \vspace{-0.5cm}
\end{figure}

In Fig.\,\ref{fig:v1-20-30} we present results for $v_1(p_T)$ compared to experimental data from ALICE \cite{Aamodt:2011by}, extracted in \cite{Retinskaya:2012ky}, and from
ATLAS \cite{Jia:2012hx}. $v_1(p_T)$ cannot be positive definite because momentum conservation requires $\langle v_1(p_T) p_T \rangle=0$. 
There is a disagreement between the experimental results (discussed in \cite{Jia:2012hx}) 
and between theory and experiment at LHC. On the other hand, $v_1(p_T)$ at RHIC is very well reproduced (see Fig.\,\ref{fig:vn30-40-RHIC}).
One possible explanation for the data crossing $v_1(p_T)=0$ at a lower $p_T$ than the calculation at LHC could be the underestimation of the pion $p_T$-spectra at very low 
$p_T$ -- see Fig.\,\ref{fig:pt-}. However, this is not necessarily the only explanation. In fact, for RHIC energies, calculated pion spectra also underestimate the data for $p_T<300\,{\rm MeV}$ but $v_1(p_T)$ is well reproduced. 

\begin{figure}[htb]
   \begin{center}
     \includegraphics[width=8.75cm]{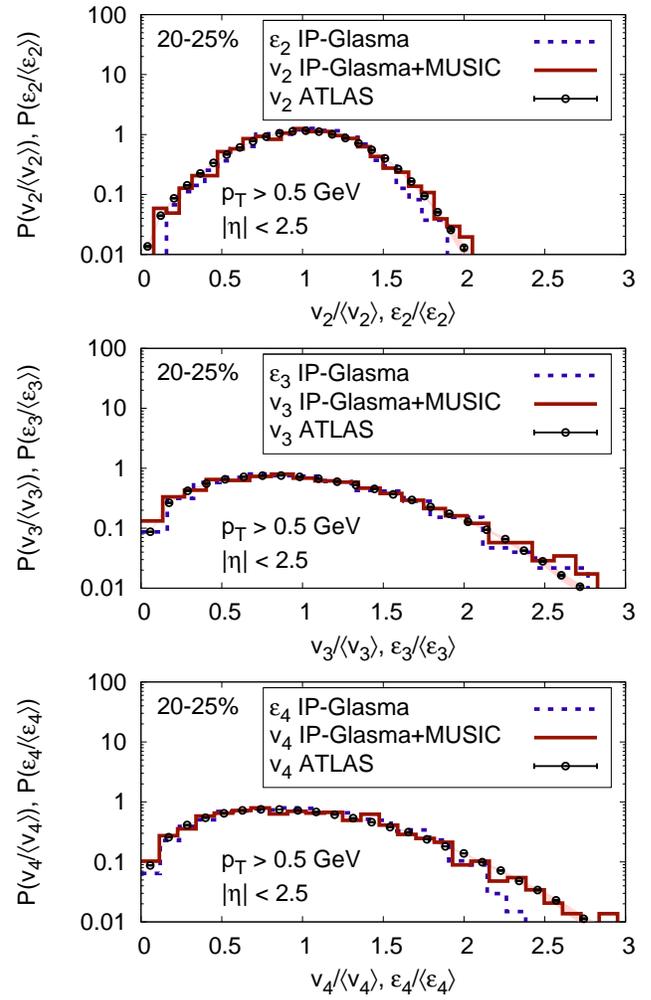}
     \vspace{-0.75cm}
     \caption{(Color online) Scaled distributions of $v_2$, $v_3$, and $v_4$ (from top to bottom) compared to experimental data from the ATLAS
     collaboration \cite{ATLAS:2012Jia,Jia:2012ve}. 1300 events. Bands are systematic experimental errors.}
     \label{fig:vnenDist-20-25}
   \end{center}
   \vspace{-0.5cm}
\end{figure}

We present event-by-event distributions of $v_2$, $v_3$, and $v_4$ compared to results from the ATLAS collaboration \cite{ATLAS:2012Jia,Jia:2012ve} 
in Fig.\,\ref{fig:vnenDist-20-25}. We chose 20-25\% central events because eccentricity distributions from neither MC-Glauber nor MC-KLN models
agree with the experimental data in this bin \cite{Jia:2012ve}.
To compare data with the distribution of initial eccentricities
\footnote{We define $\varepsilon_n = \sqrt{\langle r^n \cos(n\phi)\rangle^2+\langle r^n \sin(n\phi)\rangle^2}/{\langle r^n \rangle}$, where ${\langle \cdot \rangle}$ is the energy density weighted average.} 
from the IP-Glasma model and the final $v_n$ distributions after hydrodynamic evolution, we scaled the distributions by their respective mean value.
We find that the initial eccentricity distributions are a good approximation to the distribution of experimental $v_n$. Only for $v_4$ 
(and less so for $v_2$) the large $v_n$ end of the experimental distribution is much better described by the hydrodynamic $v_n$ distribution 
than the $\varepsilon_n$ distribution. This can be explained by non-linear mode coupling becoming important for large values of $v_2$ and $v_4$.

In summary, we have shown that the IP-Glasma+\textsc{music} model gives very good agreement to multiplicity and flow distributions at RHIC and LHC. By including properly sub-nucleon scale color charge fluctuations and their resulting early time CYM dynamics, this model significantly extends previous studies in the literature \cite{Luzum:2008cw,Takahashi:2009na,Holopainen:2010gz,Song:2011qa,Qiu:2011iv,Schenke:2011bn,Bozek:2011ua}. Omitted in all studies including ours is the stated dynamics of instabilities and strong scattering in over-occupied classical fields that can drive the system to isotropy and generate substantial flow well prior to thermalization. Ongoing work in this direction is promising and can be incorporated seamlessly in our framework. In addition, there are uncertainties in the equation of state, and in chemical and thermal freeze-out assumptions and parameters. We have not attempted a fine tuning of parameters -- the sensitivity of our results to various parameters will be addressed in a follow up work. Despite these caveats, the successful description of a wide range of data in our model provides a framework to nail down key aspects of the complex dynamics of heavy ion collisions.

\emph{Acknowledgments}
BPS\ and RV\ are supported under DOE Contract No.DE-AC02-98CH10886 and acknowledge additional support from a BNL Lab Directed Research and Development grant.
CG and SJ are supported by the Natural Sciences and Engineering Research Council of Canada.
We gratefully acknowledge computer time on the Guillimin cluster at the CLUMEQ HPC centre, a part of Compute Canada HPC facilities.
BPS gratefully acknowledges a Goldhaber Distinguished Fellowship from Brookhaven Science Associates. BPS thanks G. Denicol and J. Jia for helpful discussions.

\vspace{-0.5cm}
\bibliography{spires}

\end{document}